\documentclass[aoas,nameyear,noautosecdot,dvips]{arximspdf}
\usepackage{dcolumn}
\usepackage{graphicx}

\doi{10.1214/09-AOAS262}
\volume{3}
\issue{4}
\pubyear{2009}
\firstpage{1634}
\lastpage{1654}

\makeatletter

\newcolumntype{d}[1]{D{.}{.}{#1}}

\newtheorem{theorem}{Theorem}[section]

\newproclaim{definition}{Definition}[section]

\makeatother

\begin{document}
\begin{frontmatter}

\title{Inference on low-rank data matrices with applications to microarray data\thanksref{T1}}
\runtitle{Inference on low-rank data matrices}
\thankstext{T1}{Supported in part by the NSF Grants DMS-06-04229,
DMS-08-00631, NIH Grant R01GM080503-01A1 in the US, NNSF of China Grant 10828102, and a Changjiang Visiting Professorship at
the Northeast Normal University, China.}
\begin{aug}
\author[A]{\fnms{Xingdong} \snm{Feng}\ead[label=e1]{xfeng@niss.org}} \and
\author[B]{\fnms{Xuming} \snm{He}\corref{}\ead[label=e2]{x-he@illinois.edu}}
\runauthor{X. Feng and X. He}
\affiliation{University of Illinois at Urbana-Champaign}
\address[A]{X. Feng\\
National Institute of\\
\quad Statistical Sciences\\
19 T.W. Alexander Drive\\
Durham, North Carolina 27709\\
USA\\
 \printead{e1}}
\address[B]{X. He\\
Department of Statistics\\
University of Illinois \\
\quad at Urbana-Champaign\\
725 South Wright Street\\
Champaign, Illinois 61820\\
USA\\
\printead{e2}} 

\end{aug}

\received{\smonth{8} \syear{2008}}
\revised{\smonth{4} \syear{2009}}

\begin{abstract}
Probe-level microarray data are usually stored in matrices, where
the row and column correspond to array and probe, respectively.
Scientists routinely summarize each array by a single index as the
expression level of each probe set (gene). We examine the adequacy
of a unidimensional summary for characterizing the data matrix of
each probe set. To do so, we propose a low-rank matrix model for the
probe-level intensities, and develop a useful framework for testing
the adequacy of unidimensionality against targeted alternatives.
This is an interesting statistical problem where inference has to be
made based on one data matrix whose entries are not i.i.d. We analyze the asymptotic properties
of the proposed test statistics, and use Monte Carlo simulations to assess their small
sample performance. Applications of the proposed tests to GeneChip
data show that evidence against a unidimensional model is often
indicative of practically relevant features of a probe set.
\end{abstract}


\end{frontmatter}

\section{\texorpdfstring{Introduction.}{Introduction}}\label{sec1}

Oligonucleotide expression array technology is popular in many
fields of biomedical research. The technology makes it possible to
measure the abundance of messenger ribonucleic acid (mRNA)
transcripts for a large number of genes simultaneously. One of them
is the Genechip microarray technology, which is commercially
developed by Affymetrix to measure gene expression by hybridizing
the sample mRNA on a probe set, typically composed of 11--20 pairs of
probes, in a specially designed chip that is called a ``microarray''
[Parmigiani et al. (\citeyear{Parmigiani})].

Two types of probes are used in the Genechip microarray technology,
the perfect match ($\mathit{PM}$), which is taken from a gene sequence for
specific binding of mRNA for the gene, and the mismatch ($\mathit{MM}$),
which is artificially created by changing one nucleotide of the $\mathit{PM}$
sequence to control nonspecific binding of mRNA from the other genes
or noncoding sequences of DNA. The probe pairs are immobilized into
an array, where each spot of the array contains a probe. An RNA
sample labeled with a fluorescent dye is hybridized to a microarray,
and the array are then scanned. The expression levels of different
genes can be measured by the intensities of the spots. We use $\mathit{PM}$ or
$\mathit{PM}$--$\mathit{MM}$
as the intensity data for our
statistical analysis. Extensive studies have been carried out on how
to summarize the gene expression levels based on the probe level
data. Li and Wong (\citeyear{LiAndWong}) proposed a multiplicative model:
\begin{equation}\label{LiandWong}
y_{ij}=\theta_i\phi_j +\varepsilon_{ij},\qquad
i=1,\ldots,n, j=1,\ldots,m,
\end{equation}
where $y$ is the observed intensity of each spot, $\theta$ is the
array effect, $\phi$ is the probe effect, $\varepsilon$ is the
random error, $i$ indicates the $i$th array and $j$ refers to the
$j$th probe. This model, along with some of its variations, has been
routinely used in microarray data analysis. In the present paper we
focus on one natural question: how well can we use one quantity
$\theta_i$ to adequately summarize the expression level for each
probe set in the $i$th array? Hu, Wright and Zou (\citeyear{HuAndWright}) show that the least
squares estimate (LSE) of the parameters in the model can be
obtained as the first component of the singular value decomposition
(SVD) of the intensity matrix $\mathbf{Y}$, where
\[
\mathbf{Y}=
\pmatrix{
y_{11}&\cdots&y_{1m}\cr
\vdots&\vdots&\vdots\cr
y_{n1}&\cdots&y_{nm}
}.
\]
Motivated by their work, we aim to develop useful methods to test if
additional parameters are needed to characterize the expression data
of each probe set in each array based on the SVD.

\begin{figure*}

\includegraphics{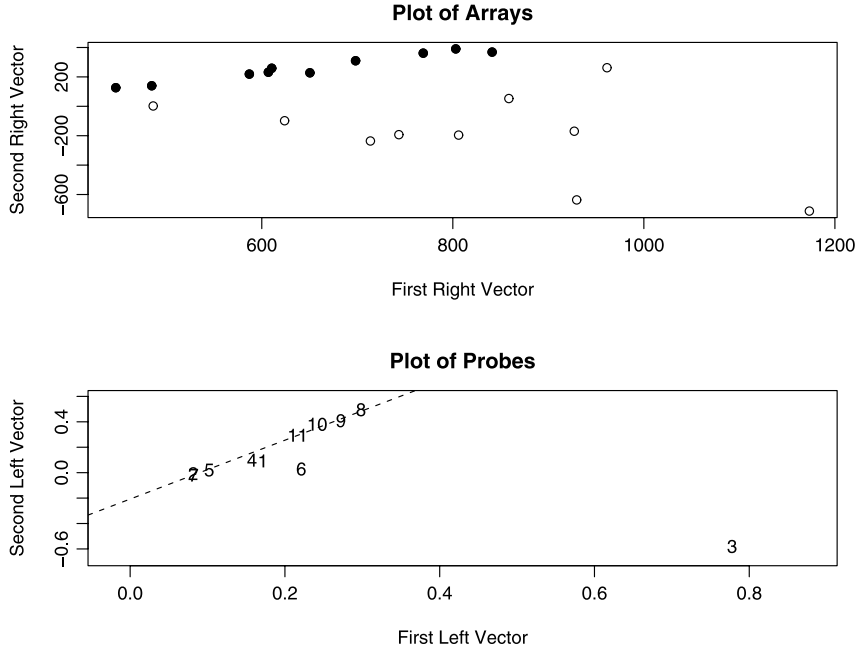}

   \caption{Scatterplot of
  singular vectors for  the probe set ``214974\_x\_at.''
  The probe numbers are shown in the lower plot, and the dotted
  line is given by the least trimmed squares estimate.
  The circles in the upper plot represent the arrays hybridized by the
  samples from the colorectal adenocarcinomas, while the
   solid points represent the arrays hybridized by the samples from the
   normal colonic tissues. Sections~\protect\ref{sec1}
   and~\protect\ref{application}
   refer to this figure.}\label{gene24270}
\end{figure*}

When we applied the SVD to the 20 GeneChip microarrays produced in a recent
MicroArray Quality Control (MAQC) project [Shi et al. (\citeyear{MAQC})] for contrasting colorectal
adenocarcinomas and matched normal colonic tissues, we found a number of
probe sets (including Probe set ``214974\_x\_at'' designed to measure the gene expression for
Gene ``CXCL5'') with a significant 2-dimensional structure. The first two singular vectors
for Probe set ``214974\_x\_at'' are displayed graphically in Figure~\ref{gene24270}, indicating
that the usual unidimensional summary of gene expression (corresponding to the first
right singular vector) would mask the differential expression of Gene ``CXCL5'' in the tumor tissues.
Recent studies, such as that reported in Dimberg et al. (\citeyear{Dimberg}), show that this gene indeed plays an important
role in colorectal cancer. More detailed findings about this probe set can be found in
 Section~\ref{application} together with additional examples.

In Section~\ref{model} we propose a 2-dimensional model to take into account both
the mean structure and the variance structure of the data matrix. We
use a multiplicative model extended from Model (\ref{LiandWong}),
but the array effects are assumed to be random, in consistency with the
fact that the arrays are typically drawn
from a larger population. The LSE of the parameters in the model can
be efficiently estimated via SVD. We are interested in the
dimensionality of the mean of this data matrix, but first we need to define it in a
precise way.
\begin{definition}\label{definition1}
Given an $n\times m$ random matrix $\mathbf{Y}$,
we define the mean matrix as $E(\mathbf{Y})$. If the rank of
$E(\mathbf{Y})$ is $k$, then the dimensionality of $\mathbf{Y}$ is
defined as $k$, where $k\in\{1,2,\ldots,\min(n,m)\}$.
\end{definition}

If the rank of $E(\mathbf{Y})$ is $k$, it is well known that the SVD
of $E(\mathbf{Y})$ has $k$ nonzero singular values, and
$E(\mathbf{Y})$ can be decomposed as
$\sum_{i=1}^k\lambda_i\underline{u}_i\underline{v}_i^T$, where
$\lambda_1\geq\lambda_2\geq\cdots\geq\lambda_k$ are the singular
values, $\underline{u}_i\in \mathbb{R}^n$ is the $i$th left vector
and $\underline{v}_i\in \mathbb{R}^m$ is the $i$th right vector, for
$i=1,2,\ldots,k$. Moreover,
\[
\underline{u}_i^T\underline{u}_j=\underline{v}_i^T\underline{v}_j=
\cases{1,&\quad $i=j$,\cr
0,&\quad $i\neq j$.
}
\]
Our primary question is whether the dimensionality (rank) of the matrix $E(\mathbf{Y})$ is one or two.
For this purpose, we formulate our hypothesis as $H_0{}\dvtx{}E(\mathbf{Y})=\lambda_1\underline{u}_1\underline{v}_1^T$
versus $H_1{}\dvtx{}E(\mathbf{Y})=\lambda_1\underline{u}_1\underline{v}_1^T+\lambda_2\underline{u}_2\underline{v}_2^T$.
It is possible to consider higher ranks of the mean matrix, but our approach
is best illustrated with the rank 2 alternative, which is also the most relevant
scenario in many applications.
In Section~\ref{test} three test statistics are proposed for this problem and their asymptotic results are given.
The asymptotic analysis based on the SVD of $\mathbf{Y}$ differs from the classical literature on the eigenvalues
and eigenvectors of a sample covariance matrix, because the latter works on a data matrix with
its mean removed, but our focus is directly on the mean of the data matrix.

When the number of microarrays in an experiment is small
due to the cost concerns, the asymptotic
distributions of the statistics proposed in Section~\ref{test}  may
not be sufficiently close to their exact distributions. Hence, we
apply the bootstrap techniques to calibrate the first two tests
discussed in Section~\ref{test}. In Section~\ref{simulation} we
 assess the finite sample performance of the
tests proposed in Section~\ref{test} by Monte Carlo simulations.
Finally, in Section~\ref{application} we
apply the proposed tests to real data sets from two studies. Our analysis shows
that the second dimension of the probe-level data is often indicative of
interesting features of a probe set. A number of scenarios for the inadequacy of a uni-dimensional
summary are discussed through the
case studies and in the concluding Section~\ref{conclude}. For example,
we point out how our approach relates to and
differs from probe remapping, and show that a high percentage of probes of poor binding strengths
in a probe set can mask gene expression profiles through a unidimensional model.
All the proofs of lemmas and theorems given in the paper can be found in the supplemental article Feng and He (\citeyear{Feng}).

\section{\texorpdfstring{Model and estimation.}{Model and estimation}}\label{model}

In this section we propose a multiplicative model
extended from Model (\ref{LiandWong}) to account for a possible
second dimension in the data matrices. Furthermore, the asymptotic
properties of the LSE of the parameters in the model are discussed.

\subsection{\texorpdfstring{A Multiplicative model with random effects.}{A Multiplicative model with random effects}}

Our proposed model takes the form
\begin{equation}\label{equ0}
\underline{y}_i=\theta_{1i}^{(0)}\underline{\phi}_1^{(0)}+\theta_{2i}^{(0)}\underline{\phi}_2^{(0)}
+\underline{\varepsilon}_i,\qquad
 i=1,2,\ldots,n,
\end{equation}
where $\underline{y}_i=(y_{i1},y_{i2},\ldots,y_{im})^T$ is the $i$th
observed vector,
$\underline{\theta}_1^{(0)}=(\theta_{11}^{(0)},\ldots,\theta_{1n}^{(0)})^T$
and
$\underline{\theta}_2^{(0)}=(\theta_{21}^{(0)},\ldots,\theta_{2n}^{(0)})^T$
are used to explain the row effects, and
$\underline{\phi}_1^{(0)}=(\phi_{11}^{(0)},\ldots,\phi_{1m}^{(0)})^T$
and
$\underline{\phi}_2^{(0)}=(\phi_{21}^{(0)},\ldots,\phi_{2m}^{(0)})^T$
are used to explain the column effects in the data matrix. When
applied to the probe level microarray data, $\theta$ stands for the
array effect and $\phi$ represents the probe effect. Using
$\|\cdot\|^2$ to denote the $L_2$ norm for vectors, and
$\underline{a}\ \bot \  \underline{b}$ for orthogonality of
$\underline{a}$ and $\underline{b}$, we make the following
assumptions:
\begin{enumerate}[(M4)]
 \item[(M1)]$\underline{\phi}_1^{(0)}$ and $\underline{\phi}_2^{(0)}$ are two $m$-dimensional unit vectors with $\underline{\phi}_1^{(0)}\, \bot \
 \underline{\phi}_2^{(0)}$.
\item[(M2)] $\underline{\theta}_{j}^{(0)}$ are
independently distributed with mean
$\underline{\mu}_j=(\mu_{j1},\ldots,\mu_{jn})^T$ and variance
$\sigma_j^2I_n$, for $j=1,2$, and all the components in each vector
are independent. The third and fourth central moments of
$\theta_{ji}^{(0)}$ are $\gamma^3_{j}$ and $\tau^4_j$, respectively,
for $j=1,2$. Moreover, $\underline{\mu}_1\, \bot \ \underline{\mu}_2$.
\item[(M3)]The error variables $\underline{\varepsilon}_i=(\varepsilon_{i1},\ldots,\varepsilon_{im})^T$
are identically and independently distributed with mean zero and
variance-covariance matrix $\sigma^2I_m$, and the third and fourth
central moments of $\varepsilon_{ij}$ are $\gamma^3$ and $\tau^4$,
respectively.
\item[(M4)]$\{\theta_{1i}^{(0)}\}$,
$\{\theta_{2i}^{(0)}\}$ and $\{\underline{\varepsilon}_i\}$ are
mutually independent.
\item[(M5)]$n^{-1}\|\underline{\mu}_{1}\|^2\rightarrow\mu_1^2$
 and
 ${n}^{-1}\|\underline{\mu}_{2}\|^2\rightarrow\mu_2^2$ as $n\rightarrow \infty$ for some finite constants $\mu_1$ and $\mu_2$. We assume that
 $\mu_1^2+\sigma^2_1>\mu_2^2+\sigma_2^2$, which is necessary for the
 identifiability of the model parameters.
\item[(M6)]$\|\underline{\mu}_j\odot\underline{\mu}_j\|^2=O(n)$, $j=1,2$, where $\odot$ indicates the pointwise product of two vectors.
 \end{enumerate}

\subsection{\texorpdfstring{Least squares estimate of column effect parameters.}{Least squares estimate of column effect parameters}}
In this section we discuss the properties of the LSE of the column
effect parameters. Let
$\underline{\theta}_1=(\theta_{11},\ldots,\theta_{1n})^T$,
$\underline{\theta}_2=(\theta_{21},\ldots,\theta_{2n})^T$,
${\varphi}=(\underline{\phi}_1^T,\underline{\phi}_2^T)^T$ and
$\vartheta=(\underline{\theta}_1^T,\underline{\theta}_2^T,\varphi^T)^T$.
With the objective function
\begin{equation}
 d_n({\vartheta})=\sum_{i=1}^n
\|\underline{y}_i-\theta_{1i}\underline{\phi}_1-\theta_{2i}\underline{\phi}_2\|^2,
\label{equ1}
\end{equation}
the least squares estimate of ${\vartheta}$ can be found by
minimizing $d_n(\vartheta)$.
In the present framework, the total number of parameters increases with
the number of observations. To facilitate the analysis, it helps to view
$\underline{\theta}_{1}^{(0)}$ and $\underline{\theta}_{2}^{(0)}$ as nuisance parameters.
If (\ref{equ1}) is minimized at
$\hat{\vartheta}$, then
$\hat{\theta}_{1i}$ and $\hat{\theta}_{2i}$ minimize
\begin{eqnarray*}
\|\underline{y}_i-\theta_{1i}\underline{\hat{\phi}}_1-\theta_{2i}\underline{\hat{\phi}}_2\|^2
\end{eqnarray*}
with respect to $\theta_{1i}$ and $\theta_{2i}$ given
$\underline{\hat{\phi}}_1$ and $\underline{\hat{\phi}}_2$.
Furthermore,
\begin{equation}\label{related1}
\underline{\hat{\theta}}_{1}=(\underline{\hat{\phi}}_1^T\underline{\hat{\phi}}_1)^{-1}\mathbf{Y}\underline{\hat{\phi}}_1,
\end{equation}
and
\begin{equation}\label{related2}
\underline{\hat{\theta}}_{2}=(\underline{\hat{\phi}}_2^T\underline{\hat{\phi}}_2)^{-1}\mathbf{Y}\underline{\hat{\phi}}_2.
\end{equation}
Therefore, $\hat \varphi$ minimizes the following objective function:
 \begin{equation}
d^*_n (\varphi)=\sum_{i=1}^n
\|\underline{y}_i-[(\underline{\phi}_1^T\underline{\phi}_1)^{-1}\underline{\phi}_1^T\underline{y}_i]\underline{\phi}_1-[(\underline{\phi}_2^T\underline{\phi}_2)^{-1}\underline{\phi}_2^T\underline{y}_i]\underline{\phi}_2\|^2.
\label{equ2}
\end{equation}

\subsubsection{\texorpdfstring{Consistency and asymptotic representation.}{Consistency and asymptotic representation}}
We consider the asymptotic properties of $\hat \varphi$ assuming that the number of probes $m$ is
fixed but the number of arrays $n \to \infty$. As shown in the preceding subsection, $\hat \varphi$ is
a constrained M estimator that minimizes (\ref{equ2}) subject to
$\|\underline{{\phi}}_1\|=\|\underline{{\phi}}_2\|=1$ and $\underline{\phi}_1\ \bot \ \underline{\phi}_2$.
The derivations in the Appendix lead to the following results.
\begin{theorem}\label{lemma2}
 When Model (\ref{equ0}) and assumptions \textup{(M1)--(M6)}
hold, $\hat{{\varphi}}\stackrel{a.s.}{\longrightarrow}{\varphi}^{(0)}$,
where $\hat{{\varphi}}$ is the least squares estimate of
${\varphi}^{(0)}$, that is, $\hat{\varphi}$ minimizes
$\sum_{i=1}^n\rho(\underline{y}_i; \varphi)$ subject to
$\|\underline{\phi}_1\|=\|\underline{\phi}_2\|=1$ and $\underline{\phi}_1\, \bot \ \underline{\phi}_2$, where
\begin{equation}
\rho(\underline{y}_i;\varphi)=
\|\underline{y}_i-(\underline{\phi}_1^T\underline{y}_i)\underline{\phi}_1-(\underline{\phi}_2^T\underline{y}_i)
\underline{\phi}_2\|^2.\label{equ3}
 \end{equation}
\end{theorem}

Theorem~\ref{lemma2} makes it possible for us to give the Bahadur
representation for $\underline{\hat{\phi}}_1$ and
$\underline{\hat{\phi}}_2$ from the results of He and Shao (\citeyear{HeAndShao}). We
now consider the limiting distribution of
$\sqrt{n}(\hat{\varphi}-\varphi^{(0)})$, which is critical for us to
discuss the asymptotic properties of the test statistics proposed in
Section~\ref{test}. Let
 \begin{equation}\label{Gamma_n}
\hspace*{10pt}\Gamma_n= (n^{-1}\|\underline{\mu}_{1}\|^2+\sigma_1^2 )\underline{\phi}_1^{(0)}
\underline{\phi}_1^{(0)T}+ (n^{-1}\|\underline{\mu}_{2}\|^2+\sigma_2^2 )
\underline{\phi}_2^{(0)}\underline{\phi}_2^{(0)T}+\sigma^2I_m,
\end{equation}
where $I_m$ is an $m\times m$ identity matrix.
Then we have the
following theorem.
\begin{theorem}\label{normal}
 When Model \textup{(\ref{equ0})} and Assumptions \textup{(M1)--(M6)}
hold, we have, for $j=1,2$,
\begin{eqnarray}\label{equPhi1}
\underline{\hat{\phi}}_j-\underline{\phi}^{(0)}_j&=&-n^{-1}D_{jn}^{-1}\sum _{i=1}^n
\bigl[2\underline{y}_i\underline{y}_i^T\underline{\phi}^{(0)}_j-2\bigl(\underline{\phi}_j^{(0)T}
\underline{y}_i\underline{y}_i^T\underline{\phi}_j^{(0)}\bigr)\underline{\phi}_j^{(0)}\bigr]\nonumber
\\[-8pt]\\[-8pt]
&&{}+o({n^{-1+\epsilon}}),\nonumber
\end{eqnarray}
where $\epsilon$ is any positive number, and
\begin{equation}
D_{jn}=-2\Gamma_n+2\underline{\phi}_j^{(0)T}\Gamma_n\underline{\phi}_j^{(0)}I_m+
4\underline{\phi}_j^{(0)}\underline{\phi}_j^{(0)T}\Gamma_n.\label{equBn}
\end{equation}
\end{theorem}

Thus, both $\sqrt{n}(\underline{\hat{\phi}}_1-\underline{\phi}_1^{(0)})$ and
$\sqrt{n}(\underline{\hat{\phi}}_2-\underline{\phi}_2^{(0)})$
are asymptotically normally distributed with mean 0 and variance-covariance
matrix, say, $C_1$ and $C_2$, respectively, where $C_1$ and $C_2$ are
determined by $\varphi^{(0)}$ and the first four moments of $\underline{y}_i$.

\subsection{\texorpdfstring{Least squares prediction of row effects.}{Least squares prediction of row effects}}

We now discuss the asymptotic properties of the least
squares prediction of the row effects based on (\ref{related1}) and
(\ref{related2}). The result is summarized in the following theorem.
\begin{theorem}\label{pred}
 When Model (\ref{equ0}) and assumptions \textup{(M1)--(M6)}
hold, we have
$\hat{\theta}_{1i}=\underline{\hat{\phi}}_1^T\underline{y}_i\stackrel{L}{\longrightarrow}
\theta_{1i}^{(0)}+\underline{\varepsilon}_i^T\underline{\phi}_1^{(0)}$
and
$\hat{\theta}_{2i}=\underline{\hat{\phi}}_2^T\underline{y}_i\stackrel{L}{\longrightarrow}
\theta_{2i}^{(0)}+\underline{\varepsilon}_i^T\underline{\phi}_2^{(0)}$,
where $\stackrel{L}{\longrightarrow}$ denotes convergence in distribution.
\end{theorem}

Let
\begin{equation}\label{Gamma}
\Gamma=(\mu^2_1+\sigma_1^2)\underline{\phi}_1^{(0)}
\underline{\phi}_1^{(0)T}+(\mu^2_2+\sigma_2^2)\underline{\phi}_2^{(0)}\underline{\phi}_2^{(0)T}+\sigma^2I_m.
\end{equation}
The first two eigenvalues of this matrix are $\mu_1^2 + \sigma_1^2 +\sigma^2$ and $\mu_2^2 + \sigma_2^2 +\sigma^2$,
with the remaining eigenvalues $\sigma^2$. Let
\begin{eqnarray}\label{equSn}
S_n=n^{-1}\mathbf{Y}^T\mathbf{Y}-n^{-1}\|\mathbf{Y}\hat{\underline{\phi}}_1\|^2-
n^{-1}\|\mathbf{Y}\hat{\underline{\phi}}_2\|^2.
\end{eqnarray}
Then, from
(\ref{related1}), (\ref{related2})  and Theorem
\ref{normal}, we have
\begin{eqnarray*}
n^{-1}\|\underline{\hat{\theta}}_{j}\|^2\stackrel{\mathrm{a.s.}}{\longrightarrow}\mu_j^2+\sigma_j^2+\sigma^2
\qquad(j=1,2),
\end{eqnarray*}
and
\begin{eqnarray*}
{(m-2)^{-1}}{S_n}\stackrel{\mathrm{a.s.}}{\longrightarrow}\sigma^2,
\end{eqnarray*}
 based on the strong law of large numbers. These consistent estimators
 for all the eigenvalues of
 the matrix $\Gamma$ will be used when we construct the tests in the following
 section. On the other hand, we note that $\theta_{1i}^{(0)}$ and $\theta_{2i}^{(0)}$
 may have their individual means $\mu_{1i}$ and $\mu_{2i}$,
 respectively, and, thus, it is impossible to consistently estimate the individual
  parameters $\mu_{1i}$, $\mu_{2i}$, $\sigma_1^2$ and $\sigma_2^2$ without any further information.

\section{\texorpdfstring{Hypothesis testing.}{Hypothesis testing}}\label{test}
 In this section we consider testing the null
hypothesis that $H_0{}\dvtx{}\underline{\mu}_2=\underline{0}$. The second
dimension $\underline{\theta}_{2}^{(0)}\underline{\phi}_2^{(0)T}$ in
Model (\ref{equ0}) does not provide meaningful information on the
mean structure of the data matrix under this null hypothesis. We
expect $\underline{\hat{\theta}}_2$ to have zero mean under the null
hypothesis and nonzero mean under the alternative hypothesis,
because
$\hat{\theta}_{2i}\stackrel{L}{\rightarrow}\theta_{2i}^{(0)}+\underline{\varepsilon}_i^T\underline{\phi}_2^{(0)}$
as $n\rightarrow\infty$. Motivated by this, we construct test
statistics based on $\{\hat{\theta}_{2i},i=1,2,\ldots,n\}$. We
consider three specific test statistics in the following
sub-sections.

\subsection{\texorpdfstring{Test on a target direction.}{Test on a target direction}}\label{average}
 Consider
\begin{equation}
T_{\underline{a}}=n^{-1}\underline{a}^T\underline{\hat{\theta}}_2,
\end{equation}
for any $\underline{a}=(a_1,\ldots,a_n)^T\in \mathbb{R}^n$ such that
$\underline{a}^T\underline{\mu}_1=0$, $\|\underline{a}\|^2=n$ and
$\max_{1\leq j\leq n}a_j^2/\break n\rightarrow 0$. We choose a vector
$\underline{a}$ such that $\underline{a}\ \bot \ \underline{\mu}_1$
because $\underline{\mu}_1$ is orthogonal to $\underline{\mu}_2$ and
we want to test the null hypothesis that $\underline{\mu}_2=0$.
We use $\underline{1}_n$ to indicate the $n$-dimensional vector with
all the components equal to 1. From
the asymptotic properties discussed in Section~\ref{model}, we have
the following theorem.
\begin{theorem}\label{tm3.1}
If the observations
$\underline{y}_1,\underline{y}_2,\ldots,\underline{y}_n$ are drawn
from Model (\ref{equ0}) and assumptions \textup{(M1)--(M6)} hold, and
$\underline{a}\in\mathbb{R}^n$ is a vector satisfying
$\underline{a}^T\underline{\mu}_1=0$,
$\underline{a}^T\underline{a}=n$ and $\max_{1\leq j\leq
n}a_j^2/n\rightarrow 0$, then
\begin{eqnarray*}
{n^{-1/2}}\underline{a}^T\underline{\hat{\theta}}_2/\hat{\sigma}
\stackrel{L}{\rightarrow}N(0,1)
\end{eqnarray*}
 under the null hypothesis that
$\underline{\mu}_2=\underline{0}$, where
\begin{equation}\label{equ310}
\hat{\sigma}^2={n^{-1}\|\underline{\hat{\theta}}_{2}\|^2-\hat{\theta}_{2\cdot}^2} \quad\mbox{and}\quad
\hat{\theta}_{2\cdot}=n^{-1}\underline{\hat{\theta}}_{2}^T\underline{1}_n.
\end{equation}
\end{theorem}

The power of the test depends on how far
$\underline{a}^T\underline{\mu}_2$ deviates from zero. As to the
target direction $\underline{a}$, it is usually determined by some
specific comparison in practice. We will give examples of choosing
$\underline{a}$ in Section~\ref{application}.

\subsubsection{\texorpdfstring{A practical solution when $\underline{\mu}_1$ is unknown.}{A practical solution when $\underline{\mu}_1$ is unknown}}
In practice, the true value of the mean vector $\underline{\mu}_1$
is unknown, but it can be estimated when extra
group information is available. Assume that the observations can be
divided into $p$ groups such that $\mu_{1i}$ are equal within each
group. We assume that $\mu_{1,n_{t-1}+1}=\cdots=\mu_{1n_t}$, for
$t=1,2,\ldots,p$, where $n_0=0< n_1<\cdots <n_{p-1}< n_p=n$, and
assume that $p$ is fixed but $n_t-n_{t-1}\rightarrow\infty$ when
$n\rightarrow \infty$. For microarray data, those arrays that use the
same types of tissues may form one group, and specific examples will be
discussed in Section~\ref{application}.

Suppose that $\hat{\mu}_{1n_t}$ is a consistent estimator of $\mu_{1n_t}$
Let
\begin{eqnarray*}
\underline{\hat{\mu}}_1=(\hat{\mu}_{1n_1},\ldots,\hat{\mu}_{1n_1},\hat{\mu}_{1n_2},\ldots,\hat{\mu}_{1n_2},\ldots,\hat{\mu}_{1n_p},\ldots,\hat{\mu}_{1n_p})^T,
\end{eqnarray*}
 where the number of $\hat{\mu}_{1n_t}$ in the above vector is $n_t-n_{t-1}$, $t=1,2,\ldots,p$. Furthermore, when we choose a vector $\underline{\hat{a}}$ orthogonal to
$\underline{\hat{\mu}}_1$, we only consider the candidates whose
entries can be divided into groups and are equal to each other
within each group in the form of
\begin{eqnarray*}
\underline{\hat{a}} \propto (\hat{a}_{n_1},\ldots,\hat{a}_{n_1},\hat{a}_{n_2},\ldots,\hat{a}_{n_2},
\ldots,\hat{a}_{n_p},\ldots,\hat{a}_{n_p})^T.
 \end{eqnarray*}

With $\underline{\hat{a}}$ convergent to $\underline{a}$, the statistic $T_{\underline{\hat{a}}}=n^{-1}\underline{\hat{a}}^T\underline{\hat\theta}_2$ has the same Bahadur
representation as if we chose a vector $\underline{a}$ orthogonal to
$\underline{\mu}_1$ under the null hypothesis. Hence, when we
construct the tests in Section~\ref{test}, we can use
$\underline{\hat{a}}$ that is orthogonal to
$\underline{\hat{\mu}}_1$. The choice of $\underline{\hat{a}}$ is not unique, and is best chosen
in response to specific alternatives of interest in a given experiment.

\subsection{\texorpdfstring{A $\chi^2$ test with multiple directions.}{A $\chi^2$ test with multiple directions}} \label{chi2}

As shown in Section~\ref{average}, the power of
the test $T_{\underline{a}}$ depends on the direction
$\underline{a}$ that we choose. In some cases, we may
consider several directions simultaneously. Let us consider a
$k\times n$ matrix $A$, where $k$ is a fixed integer and $k< n$. The
$i$th row of the matrix $A$ is denoted as $\underline{a}_i$ and the
$j$th component of $\underline{a}_i$ is denoted as $a_{ij}$ for
$i=1,\ldots,k$ and $j=1,\ldots,n$. Assume that
$\underline{a}_i\ \bot \ \underline{a}_j$ for $i\neq j$,
$\underline{a}_i\ \bot \ \underline{\mu}_1$,
$\underline{a}_i^T\underline{a}_i=n$ and $\max_{1\leq j\leq
n}a_{ij}^2/n\rightarrow 0$ for each $i$. Then, we propose the
test statistic
\begin{eqnarray*}
T_A = n^{-1}\|A\underline{\hat{\theta}}_2\|^2/\hat{\sigma}^2,
\end{eqnarray*}
with the following result.
\begin{theorem}\label{theorem2}
Under the assumptions of Theorem~\ref{tm3.1}, and for the matrix $A$ described
in this subsection, we have $T_A \rightarrow \chi^2_k$ in distribution
under the null hypothesis that
$\underline{\mu}_2=\underline{0}$, where $\chi^2_k$ has the chi-square
distribution with $k$ degrees of freedom.
\end{theorem}

In practice, given observations, we should not choose $k$ that is
close to $n$, because
\begin{eqnarray*}
\|A\underline{\hat{\theta}}_2\|^2=n^2\hat{\sigma}^2
\end{eqnarray*}
when $k=n-1$, and the variations accumulated from approximation errors
will ruin the chi-square approximation.

\subsection{\texorpdfstring{Bootstrap calibration.}{Bootstrap calibration}}
Sometimes, the sample size $n$ is too small for the asymptotic
approximations to perform well. Hence, we propose a finite sample
adjustment to control the type I errors.

A bootstrap method, which avoids resampling from the rows or columns
of the data matrix, to test the null hypothesis that
$\underline{\mu}_2=0$ can be described as follows:
\begin{enumerate}[(ii)]
\item[(i)] Draw
$n$ copies $\{j_1,\ldots,j_n\}$ with replacement from
$\{1,2,\ldots,n\}$ and let
$\hat{\theta}^*_{2i}=\hat{\theta}_{2j_i}-\hat{\theta}_{2\cdot}$
$(i=1,2,\ldots,n)$, where
$\hat{\theta}_{2\cdot}=n^{-1}\sum_{i=1}^n\hat{\theta}_{2i}$, and
then evaluate $T^*_{\underline{a}}$ as
\[
T^*_{\underline{a}}={n}^{-1/2}\underline{a}^T\underline{\hat{\theta}}^*_{2}/
(n^{-1}\|\underline{\hat{\theta}}^*_{2}\|^2-\hat{\theta}^{*2}_{2\cdot})^{1/2},
\]
where
$\underline{\hat{\theta}}^*_{2}=(\hat{\theta}^*_{21},\ldots,\hat{\theta}^*_{2n})^T$
and
$\hat{\theta}^{*}_{2\cdot}=n^{-1}\sum_{i=1}^n\hat{\theta}^{*}_{2i}$;
\item[(ii)]  Repeat Step (i) for $B$
times to get the test statistic $T^*_{\underline{a},b}$,
$b=1,2,\ldots,B$. We estimate the bootstrap $p$-value by
\[
p={B}^{-1}\sum _{b=1}^{B}
I\{|T^{*}_{\underline{a},b}|\geq |T_{\underline{a}}|\}.
\]
\end{enumerate}

To see this bootstrap method work, we note that
\begin{eqnarray*}
{n}^{-1/2}\underline{a}^T\underline{\hat{\theta}}_{2\cdot}&=& \Biggl({n}^{-1/2}\sum _{i=1}^na_i\underline{\phi}_2^{(0)T}\underline{y}_i \Biggr)+o_p(1),
\\
{n}^{-1/2}\underline{a}^T\underline{\hat{\theta}}^*_{2}
&=& \Biggl(n^{-1/2}\sum _{i=1}^na_i\underline{\phi}_2^{(0)T}\underline{y}^*_i \Biggr)+o_p(1),
\end{eqnarray*}
where $\underline{y}_i^*=\underline{y}_{j_i}$, and
\begin{eqnarray*}
n^{-1}\|\underline{\hat{\theta}}_2\|^2-\hat{\theta}_{2\cdot}^2-(n^{-1}\|\underline{\hat{\theta}}_2^*\|^2-\hat{\theta}_{2\cdot}^{*2})=o_p(1).
\end{eqnarray*}
Since
\begin{eqnarray*}
 \Biggl(n^{-1}\sum _{i=1}^n\bigl(\underline{\phi}_2^{(0)T}\underline{y}_i\bigr)^2
 - \Biggl[n^{-1}\sum _{i=1}^n\underline{\phi}_2^{(0)T}\underline{y}_i \Biggr]^2 \Biggr)^{-1/2}
 \Biggl({n}^{-1/2}\sum _{i=1}^na_i\underline{\phi}_2^{(0)T}\underline{y}_i \Biggr)\stackrel{L}{\rightarrow}N(0,1)
\end{eqnarray*}
under the null hypothesis, the bootstrap method works by Theorem 1
of Mammen (\citeyear{Mammen}). Our proposed bootstrap method acts on
$\hat{\theta}_2 $, and avoids repeated computations of the SVD. The
same idea can be used for $T_A$.

\subsection{\texorpdfstring{Test based on maximum over directions.}{Test based on maximum over directions}}\label{max}
 If we do not have guided directions to look for patterns
in $\underline{\mu}_2$, we may wish to search over a larger number
of directions. The chi-square test in Section~\ref{chi2} does not
apply when $k$ is large. However, the maximum over $k=n-1$
directions,
\begin{equation}\label{Mn}
 M_n=\max _{1\leq j\leq
n-1}{n}^{-1/2}\underline{a}_{j}^T\underline{\hat{\theta}}_2,
\end{equation}
 has a simple limiting distribution when $\underline{\varepsilon}_{i}$ and $\underline{\theta}_{2}^{(0)}$ are normally distributed.
Let
\begin{equation}\label{cn}
 c_n=\sqrt{2\ln (n-1)} \quad\mbox{and}\quad
 b_n=c_n-{2}^{-1}c_n^{-1}\ln \bigl(4\pi\ln (n-1)\bigr).
\end{equation}

\begin{theorem}\label{thm3}
Assume the conditions of Theorem~\ref{tm3.1}, with the additional assumption that
$\underline{\theta}_{2}^{(0)}$ and
$\underline{\varepsilon}_{i}$ are normally distributed.
For any matrix $A$ as described
in Section~\ref{chi2} with $k=n-1$, we have
$P (c_n({M_n}/{\hat{\sigma}}-b_n)\leq x )\rightarrow
e^{-e^{-x}}$ as $n\rightarrow \infty$ under the null hypothesis that
$\underline{\mu}_2=\underline{0}$.
\end{theorem}

Under the alternative hypothesis, we should observe larger values of
$ M_n $. Furthermore, the convergence rate of the extreme statistic
is discussed in Section 4.6 of Leadbetter, Lindgren and Rootzen (\citeyear{Leadbetter}). Based on their
arguments, we can use $[\Phi(u)]^{n-1}$ to approximate the
probability $P ({M_n}/{\hat{\sigma}}\leq u )$ in computing the
$p$-values of the proposed test here.

The normality of $\underline{\theta}_{2}^{(0)}$ and
$\underline{\varepsilon}_{i}$ is not a necessary condition for the
limiting distribution to hold. Our simulation results not reported
in this paper suggest that Theorem~\ref{thm3} may hold in a much
broader setting.

\section{\texorpdfstring{Simulations.}{Simulations}}\label{simulation}

To assess the performance of the proposed tests in the present paper,
we report Monte Carlo simulation results by simulating data from Model (\ref{equ0}),
with the following specifications. The size of the parameters are
chosen to mimic some real microarray data:
\begin{enumerate}[(iii)]
\item[(i)] $\underline{\theta}_1^{(0)}$
is generated from the multivariate
$N(\underline{\mu}_1, 150{,}000I_n)$, where
$\underline{\mu}_1=(4500,4500,\ldots,4500)^T$;
\item[(ii)] $\underline{\theta}_2^{(0)}$
is generated from
$N(\underline{\mu}_2, 10{,}000I_n)$, where $\underline{\mu}_2$ is
equal to either $(0,0,\ldots,0)^T$ as the null hypothesis or
$(125,-125,\ldots,125,-125)^T$ as an alternative hypothesis;
\item[(iii)] $\underline{\phi}_1= (2\sqrt{3} )^{-1}(1,1,\ldots,1)^T$
and
$\underline{\phi}_2= (2\sqrt{3} )^{-1}(1,-1,\ldots,1,-1)^T$
are of dimension 12;
\item[(iv)] The errors ${\varepsilon}_{ij}
(i=1,2,\ldots,n,j=1,2,\ldots,12)$ are drawn from three different
distributions in different experiments: the normal distribution
$N(0,5000)$, the $t$-distribution with 5 degrees of freedom
multiplied by $10\sqrt{30}$ and the centered $\chi^2$-distribution
$50(Z^2-1)$, where $Z\sim N(0,1)$.
\end{enumerate}

\subsection{\texorpdfstring{Test on a target direction.}{Test on a target direction}}\label{sim1}
 Four different sample sizes are used: $n=8,16,32$ and
128. Furthermore, we chose two different $\underline{a}$ to
compare the performance of the tests $T_{\underline{a}}$ discussed
in Section~\ref{test}.

 \subsubsection{\texorpdfstring{Case 1.}{Case 1}}
 In the first case, we choose
 $\underline{a}=(1,-1,\ldots,1,-1)^T$, which is the ideal choice for detecting the alternative in our settings.
 We draw 5000 data sets,
and the 5000 $p$-values are calculated based on the limiting
distributions in Theorems~\ref{tm3.1}. For the test
$T_{\underline{a}}$, the type I errors are close to the nominal
level of 0.05 when $n\geq 16$. Also clear from Table~\ref{error1} is
that the power of the test is decent even when the sample size is as
small as 8.

\begin{table}
  \caption{Type I errors and powers of the
target direction test are listed with increasing sample size $n$.
The errors are generated from three different distributions}\label{error1}
\begin{tabular*}{\textwidth}{@{\extracolsep{\fill}}lcccccc@{}}
\hline
&\multicolumn{3}{c}{\textbf{Null}}
&\multicolumn{3}{c@{}}{\textbf{Alternative}}\\[-6pt]
&\multicolumn{3}{c}{\hrulefill}
&\multicolumn{3}{c@{}}{\hrulefill}\\
 \textbf{Size}
 &\textbf{Normal}
 &$\bolds{t}$
 &$\bolds{\chi}^2$
 &\textbf{Normal}
 &$\bolds{t}$
 &$\bolds{\chi}^2$\\
\hline
 \phantom{00}8\tabnoteref{ta}&0.0560&0.0510&0.0430&0.6362&0.6088&0.5718 \\
 \phantom{0}16\tabnoteref{ta}&0.0542&0.0492&0.0426&0.9540&0.9308&0.9004\\
 \phantom{0}32\tabnoteref{ta}&0.0470&0.0500&0.0460&0.9998&0.9974&0.9940 \\
 128\tabnoteref{ta}&0.0522&0.0508&0.0532& 1.0000&1.0000&1.0000\\
  \phantom{00}8\tabnoteref{tb}&0.0552&0.0568&0.0458&0.4202& 0.4104&0.3854 \\
 \phantom{0}16\tabnoteref{tb}&0.0530&0.0500&0.0490&0.8358&0.8190&0.7840\\
 \phantom{0}32\tabnoteref{tb}&0.0546&0.0494&0.0440&0.9934&0.9890&0.9854 \\
 128\tabnoteref{tb}&0.0522&0.0514&0.0486& 1.0000&0.9998&1.0000\\
 \hline
\end{tabular*}
\tabnotetext[a]{ta}{The results are from Case 1.}
\tabnotetext[b]{tb}{The results are from Case 2.}
\end{table}

\subsubsection{\texorpdfstring{Case 2.}{Case 2}}

We choose
\begin{eqnarray*}
\underline{a}=2^{-1}\sqrt{3} (1,-1,\ldots,1,-1)^T+2^{-1}(1,\ldots,1,-1,\ldots,-1)^T
\end{eqnarray*}
to see whether the test has the meaningful power when
$\underline{a}$ is not so well chosen to target the true pattern in
$\underline{\mu}_2$. The results are given in the lower half of Table~\ref{error1}. A
comparison with Case 1 shows that the power of
the test $T_{\underline{a}}$ is sensitive to the choice of
$\underline{a}$ for small $n$, so a good target direction based on
the nature of the experiment or the knowledge of the experimenter is
very valuable.

\subsection{\texorpdfstring{The $\chi^2$ test.}{The $\chi^2$ test}} \label{sim2}

For the $\chi^2$ test of Section
\ref{chi2}, four sample sizes $n=8,16,32,64$ are used with the
Monte Carlo sample size of 5000. We generated $k=4$ vectors, which
are orthogonal to $\underline{\mu}_1$, orthogonal to each other, and
are of length $n$. The algorithm to generate the vectors can be
described as follows:
\[
A=\pmatrix{1&1\cr
1&-1}
\otimes\cdots\otimes
\pmatrix{1&1\cr
1&-1},
\]
where $\otimes$ is the Kronecker product, and the product is
repeated $n$ times. After the first column of $A$ is deleted, the
next $k=4$ columns are the vectors we use in the $\chi^2$ test. The
estimated type I errors and powers of the test are listed in Table~\ref{errors9}. It is clear that the type I error is not close to
0.05 when $n \le 16$. In fact, we find that the type I errors in
Table~\ref{errors6} from the limiting distributions of
$T_{\underline{a}}$ and the $\chi^2$ tests can be too high or too
low when the sample sizes $n$ are small. The bootstrap method
manages to control the type I errors even at small samples.

\begin{table}
\caption{Type I errors and powers of the
$\chi^2$ test are listed with increasing sample size $n$. The errors
are drawn from three distributions}\label{errors9}
\begin{tabular*}{\textwidth}{@{\extracolsep{\fill}}lcccccc@{}}
\hline
&\multicolumn{3}{c}{\textbf{Null}}
&\multicolumn{3}{c@{}}{\textbf{Alternative}}\\[-6pt]
&\multicolumn{3}{c}{\hrulefill}
&\multicolumn{3}{c@{}}{\hrulefill}\\
\textbf{Size}
&\textbf{Normal}
&$\bolds{t}$
&$\bolds{\chi}^2$
&\textbf{Normal}
&$\bolds{t}$
&$\bolds{\chi}^2$\\
\hline
\phantom{0}8&0.0000&0.0000&0.0000&0.0000&0.0000&0.0000 \\
 16&0.0296&0.0264&0.0236&0.6406&0.6104&0.5734 \\
 32&0.0418&0.0384&0.0394&0.9918&0.9846&0.9822 \\
 64&0.0464&0.0504&0.0422& 1.0000&0.9990&1.0000 \\
\hline
\end{tabular*}
\end{table}

\begin{table}[b]
\caption{Type I errors and powers are listed for comparison
between the bootstrap and the large-sample approximation. The errors
are generated from three different
distributions}\label{errors6}
\begin{tabular*}{\textwidth}{@{\extracolsep{\fill}}lcccccc@{}}
\hline
&\multicolumn{3}{c}{\textbf{Asymptotic approximation}}
&\multicolumn{3}{c@{}}{\textbf{Bootstrap}}
\\[-6pt]
&\multicolumn{3}{c}{\hrulefill}
&\multicolumn{3}{c@{}}{\hrulefill}
\\
$\bolds{n}$
&\textbf{Normal}
&$\bolds{t}$
&$\bolds{\chi}^2$
&\textbf{Normal}
&$\bolds{t}$
&$\bolds{\chi}^2$
\\
\hline
\multicolumn{1}{@{}l}{Type I error}\\
 \phantom{0}6\tabnoteref{tz}&0.1174&0.1096&0.1004&0.0420&0.0416&0.0350 \\
 \phantom{0}8\tabnoteref{tz}&0.0552&0.0568&0.0458&0.0484&0.0520&0.0440\\
 \phantom{0}8\tabnoteref{tm}&0.0000&0.0000&0.0000&0.0406&0.0396&0.0256 \\
 16\tabnoteref{tm}&0.0296&0.0264&0.0236&0.0520&0.0430&0.0420\\[5pt]
\multicolumn{1}{@{}l}{Estimated power}\\
 \phantom{0}6\tabnoteref{tz}&0.4950&0.4820&0.4646&0.2560&0.2380&0.2194\\
 \phantom{0}8\tabnoteref{tz}&0.4202& 0.4104&0.3854&0.3738&0.3670&0.3480\\
 \phantom{0}8\tabnoteref{tm}&0.0000&0.0000&0.0000&0.1508&0.1506&0.1338\\
 16\tabnoteref{tm}&0.6404&  0.6104&0.5734&0.7142&0.6912&0.6746\\
 \hline
\end{tabular*}
\tabnotetext[a]{tz}{The results are from the test on the target direction
$2^{-1}\sqrt{3}(1,-1,\ldots,1,-1)^T+2^{-1}(1,\ldots,1,-1,\ldots,-1)^T$.}
\tabnotetext[b]{tm}{The results are from the $\chi^2$ test based on the four
target directions.}
\end{table}

\subsection{\texorpdfstring{Test based on maximum over directions.}{Test based on maximum over directions}}

Similar to Table~\ref{errors9},
Table~\ref{errors8} shows the performance of the test $M_n$ of
Section~\ref{max} based on the limiting distributions. The test is
conservative for small $n$, but remains quite powerful in the study.
The test can be used even when the normality assumption in Theorem
\ref{thm3} is violated. However, our simulation results that are not
reported here suggest that if $\underline{\theta}_{2}^{(0)}$ and
$\underline{\varepsilon}_{i}$ do not have finite 4th moments, the
limiting distribution would not take effect for realistic sample
sizes considered in this paper.

\begin{table}
\caption{Type I errors and powers of the test based on maximum over
directions are listed with increasing sample size $n$. The errors
are drawn from three distributions}\label{errors8}
\begin{tabular*}{\textwidth}{@{\extracolsep{\fill}}lcccccc@{}}
\hline
&\multicolumn{3}{c}{\textbf{Null}}
&\multicolumn{3}{c@{}}{\textbf{Alternative}}\\[-6pt]
&\multicolumn{3}{c}{\hrulefill}
&\multicolumn{3}{c@{}}{\hrulefill}\\
 \textbf{Size}
 &\textbf{Normal}
 &$\bolds{t}$
 &$\bolds{\chi}^2$
 &\textbf{Normal}
 &$\bolds{t}$
 &$\bolds{\chi}^2$\\
 \hline
 \phantom{0}8&0.0018&0.0012& 0.0008& 0.0270&0.0268& 0.0232 \\
 16&0.0306&0.0256&0.0190&0.6992&0.6620&0.6216 \\
 32&0.0378&0.0376&0.0264&0.9850&0.9766&0.9666 \\
 64&0.0428& 0.0404&0.0362&1.0000&0.9988&0.9994 \\
 \hline
 \end{tabular*}
 \end{table}

\section{\texorpdfstring{Case studies.}{Case studies}}\label{application}

In this section we analyze two
microarray data sets. We apply our testing methods to search for
genes with potentially complicated mean structure, and further
analyze some of those genes to understand the possible causes. The data are quantile
normalized in each case.

\subsection{\texorpdfstring{Example 1.}{Example 1}}\label{example1}

We considered the GeneChip data
(\href{http://www.ncbi.nlm.nih.gov/geo/query/acc.cgi?acc=GSE5350}{http://www.ncbi.}
 \href{http://www.ncbi.nlm.nih.gov/geo/query/acc.cgi?acc=GSE5350}{nlm.nih.gov/geo/query/acc.cgi?acc=GSE5350}) obtained from
the recent MicroArray Quality Control (MAQC)
project and used in Lin et al. (\citeyear{Guixian}). We have a total of 20
microarrays (HG-U133-Plus-2.0), generated from five colorectal
adenocarcinomas and five matched normal colonic tissues with 1 technical
replicate at each of two laboratories involved in the MAQC project.

In this study we use $\mathit{PM}$ as the intensity measure in $\textbf{Y}$, and carry out
the SVD to get the two largest singular values  $\hat{\lambda}_1 >
\hat{\lambda}_2$.
We focus on 350 probe sets with the highest ratios
$\hat{\lambda}_2^2/\hat{\lambda}_1^2 $ (with all those ratios above
$1/10$). For each probe set, the probe-level microarray data are stored in
a matrix, where the rows correspond to the probes and the columns
correspond to the arrays. The intensities from the normal tissues
are entered in the column 1--5, 11--15, and those from the tumors
entered in the rest of columns.

We choose a target direction to contrast the two groups in the study. In particular,
we use
\[
\underline{a}_1 \propto
(-\hat{\mu}_2,\ldots,-\hat{\mu}_2,\hat{\mu}_1,\ldots,\hat{\mu}_1,-\hat{\mu}_2,\ldots,
   -\hat{\mu}_2,\hat{\mu}_1,\ldots,\hat{\mu}_1)^T,
\]
where $\hat{\mu}_1$ is taken to be the
median of $\hat{\theta}_{1 i}$ of the first group (normal tissues),
and $\hat{\mu}_2$ the median of $\hat{\theta}_{1 i}$
of the other group.
Hence, we have $\underline{a}_1\, \bot \ \underline{\hat{\mu}}$, where
  \begin{eqnarray*}
  \underline{\hat{\mu}}=(\hat{\mu}_1,\ldots,\hat{\mu}_1,\hat{\mu}_2,\ldots,\hat{\mu}_2,\hat{\mu}_1,\ldots,\hat{\mu}_1,\hat{\mu}_2,\ldots,\hat{\mu}_2)^T.
  \end{eqnarray*}

By the statistical test $T_{\underline{a}}$ developed in Section
\ref{average}, we find that 81 out of 350 probe sets are detected as
individually significant at the 0.05 level. Out of those, 36
probe sets remain significant after the multiple test adjustment of
Benjamini and Hochberg (\citeyear{Benj1}).

We plot $(\hat{\theta}_{1i},\hat{\theta}_{2i})$, $i=1,2,\ldots,20$,
and $(\hat{\phi}_{1j},\hat{\phi}_{2j})$, $j=1,2,\ldots,m$, for those
probe-sets that are detected as significant; some interesting facts
can be observed. We now zoom in on three of those probe sets.

\subsubsection{\texorpdfstring{Probe set ``214974\_x\_at.''}{Probe set ``214974\_x\_at''}}
In the study the probe set ``214974\_x\_at'' is used to measure the
expression level of Gene ``CXCL5.'' Our test gave the $p$-value of
$1.11 \times 10^{-3}$, the adjusted $p$-value of $2.38 \times
10^{-2}$ and the $q$-value, as proposed in Storey (\citeyear{Storey}), of $5.77\times 10^{-4}$,
offering significant evidence against the unidimensional model.
The first four singular values of the data matrix are (3387, 1388,
361, 168). As mentioned in the Introduction with Figure~\ref{gene24270}, the arrays
cannot be easily separated by the first right singular vector, but if we
use $(\hat{\theta}_{1i},\hat{\theta}_{2i})$ jointly, the arrays are
well separated in the 2-dimensional space. The usual one-dimensional
index of the probe set is insufficient to
summarize the gene expression of ``CXCL5.''

\begin{figure*}

\includegraphics{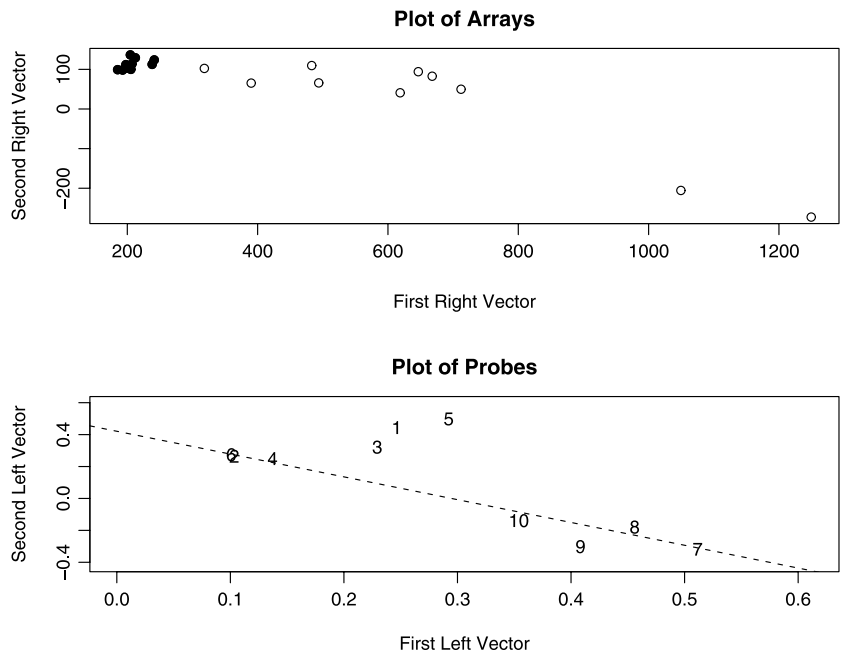}

\caption{Scatterplot of singular vectors for  the probe set ``214974\_x\_at'' after
we remove Probe 3.
  See Figure~\protect\ref{gene24270} for
   more details about this figure.}\label{gene24270rm}
\end{figure*}

Further inspection of the data shows that the intensities from Probe
3 are much higher than those of the other probes, and Probe 3
dominantly contributes to the values of $\hat{\theta}_{1i}$. By the
Basic Local Alignment Search Tool (BLAST,
\url{http://www.ncbi.nlm.nih.gov/blast/Blast.cgi}), we found that Probe~3
is represented in both Gene ``CXCL5'' and Gene ``N-PAC,'' but the
other probes were confirmed as specific to Gene ``CXCL5.'' We
further confirmed that the intensities of Probe~3 were highly
correlated with the intensities of several probes in the probe set
``208506\_x\_at'' (designed by Affymetrix to measure the expression
level of Gene ``N-PAC''), and thus, we need to take Probe 3 with
caution. If Probe 3 were removed from the probe set, we would have
seen a clear separation of the two groups from the first singular
vector; see Figure~\ref{gene24270rm}. In this case, the second
singular vector from the whole probe set appears to be a better
summary of Gene ``CXCL5.'' We note that Gene ``CXCL5'' has been
indicated as an important gene for colorectal cancer in the
literature. For example, Dimberg et al. (\citeyear{Dimberg}) observed significantly
higher expression levels of the protein encoded by ``CXCL5'' in
colorectal cancer tumors than in normal tissue, so the
multidimensionality of the probe set ``214974\_x\_at'' flagged
through our statistical work can offer biologically relevant
information.

\begin{figure*}[t]
\vspace*{-2pt}

\includegraphics{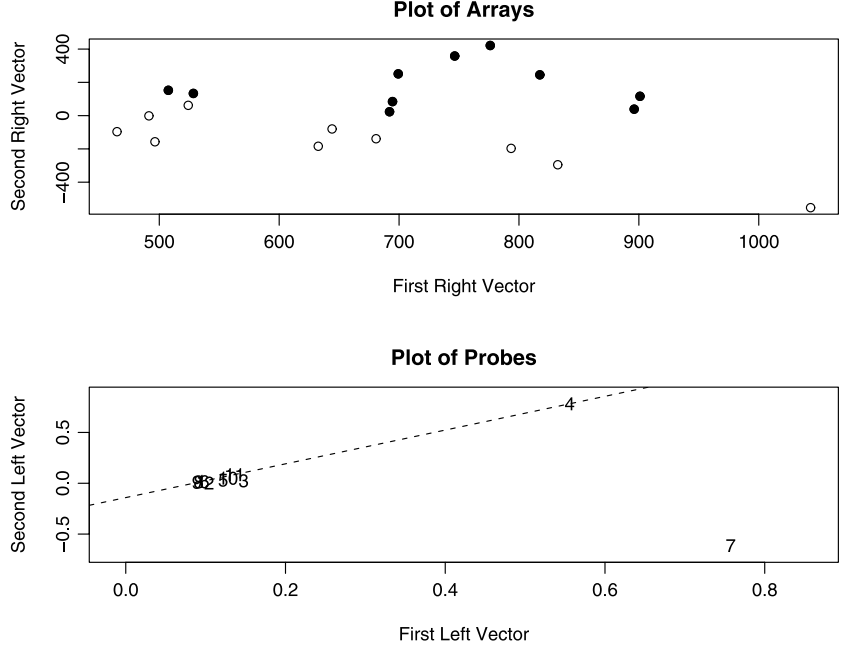}

  \caption{Scatterplot of
  singular vectors for  the probe set ``227899\_at.''
  The probe numbers are shown in the lower plot, and the dotted line is given by the
  least trimmed squares estimate.
  The circles in the upper plot represent the arrays hybridized by the samples
  from the colorectal adenocarcinomas, while the
   solid points represent the arrays hybridized by the samples from the normal colonic tissues.}\label{gene227899}
\vspace*{-2pt}
\end{figure*}

%

\subsubsection{\texorpdfstring{Probe set ``227899\_at.''}{Probe set ``227899\_at''}}
The probe set ``227899\_at'' is designed by Affymetrix to measure the
expression level of Gene ``VIT.'' Our test gave the $p$-value
$8.78\times 10^{-4}$, the adjusted $p$-value $2.38\times
10^{-2}$ and the $q$-value $5.77\times 10^{-4}$. The first four
singular values are $(3178,1011,227,77)$.

From Figure~\ref{gene227899}, we note that differential expression
can be detected from the second right singular vector, but not the
first.  From the probe-level data, we find that the intensities of
Probe 4 and Probe 7 are much higher than those of the other probes,
and these two probes dominate the first two singular vectors.
Furthermore, we confirmed by BLAST both probes as specific for
measuring the expression level of Gene ``VIT,'' and so did the other
probes. As a double check, we applied the remapping method proposed by Lu et al. (\citeyear{Lu})
and confirmed all the probes in this probe set were specified for the three transcript variants for Gene ``VIT.''
Therefore, a 2-dimensional summary of the gene appears
necessary for this probe set.

To make the point further,  we provide the absolute value of percentages calculated from $M_1$ and $M_2$
in Table~\ref{compare37154}, where
\[
M_1=
\pmatrix{
\hat{\theta}_{11}\hat{\phi}_{11}&\cdots&\hat{\theta}_{11}\hat{\phi}_{1m}\cr
\vdots&\vdots&\vdots\cr
\hat{\theta}_{1n}\hat{\phi}_{11}&\cdots&\hat{\theta}_{1n}\hat{\phi}_{1m}
}
,\]
and
\[
M_2=
\pmatrix{
\hat{\theta}_{21}\hat{\phi}_{21}&\cdots&\hat{\theta}_{21}\hat{\phi}_{2m}\cr
\vdots&\vdots&\vdots\cr
\hat{\theta}_{2n}\hat{\phi}_{21}&\cdots&\hat{\theta}_{2n}\hat{\phi}_{2m}
}.
\]

It is clear that the information contained in the second dimension for Probes 4 and 7
is important, because in more than half of the arrays their contributions from the second
dimension are more than 20\% of those from the first. The joint use of $\hat{\theta}_{1i}$ and $\hat{\theta}_{2i}$
gives a more complete picture
about the expression profile of Gene ``VIT.''

\begin{table}[b]
\vspace*{-3pt}
  \caption{A summary of the absolute values of
  $\hat{\theta}_{2i}\hat{\phi}_{2j}/\hat{\theta}_{1i}\hat{\phi}_{1j}$
  in percentage by probes}\label{compare37154}
  \begin{tabular*}{\textwidth}{@{\extracolsep{\fill}}lcd{2.2}d{2.2}d{2.2}d{2.2}@{}}
\hline
\textbf{227899\_at}
&\textbf{Min. (\%)}
&\multicolumn{1}{c}{\textbf{Q1 (\%)}}
&\multicolumn{1}{c}{\textbf{Med. (\%)}}
&\multicolumn{1}{c}{\textbf{Q3 (\%)}}
&\multicolumn{1}{c@{}}{\textbf{Max. (\%)}}\\
\hline
Probe 1   &0.06  &2.48  &5.03  &6.56 &10.92  \\
Probe 2   &0.01  &0.32 &0.64  &0.84 &1.39  \\
Probe 3   &0.04  &2.00 &4.04  &5.27 &8.78  \\
Probe 4   &0.39  &17.37 &35.20  &45.88 &76.40  \\
Probe 5   &0.07  &2.97  &6.02  &7.85 &13.06  \\
Probe 6   &0.04  &1.84  &3.72  &4.85 &8.08  \\
Probe 7   &0.22  &10.01  &20.29  &26.44 &44.02  \\
Probe 8   &0.04  &1.77  &3.59  &4.68 &7.79  \\
Probe 9   &0.03  &1.29  &2.62  &3.42 &5.69  \\
Probe 10  &0.11  &4.74  &9.61  &12.52 &20.85  \\
Probe 11  &0.17  &7.69  &15.59  &20.32 &33.83  \\
\hline
\end{tabular*}
\vspace*{-3pt}
\end{table}

\subsubsection{\texorpdfstring{Probe set ``1560296\_at.''}{Probe set ``1560296\_at''}}
The probe set ``1560296\_at'' is used in the HG-U133-Plus-2.0 platform
to represent Gene ``DST.'' This probe set is detected by our test
with a significant 2-dimensional mean structure ($p$-value $1.88\times
10^{-3}$, adjusted $p$-value $2.87\times 10^{-2}$ and $q$-value
$6.96\times 10^{-4}$). The first four singular values are $(5470,
1748,504,271)$.

\begin{figure*}

\includegraphics{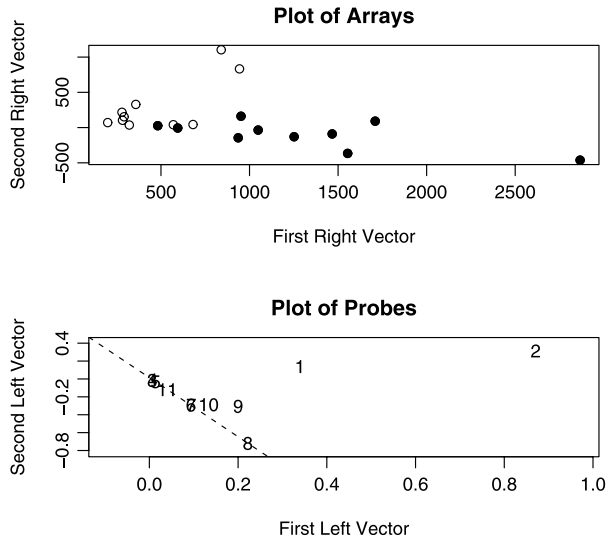}

  \caption{Scatterplot of
  singular vectors for  the probe set ``1560296\_at.''
  The probe numbers are shown in the lower plot, and the dotted
  line is given by the least trimmed squares estimate.
  The circles in the upper plot represent the arrays hybridized
  by the samples from the colorectal adenocarcinomas, while the
   solid points represent the arrays hybridized by the samples from the normal colonic tissues.}\label{gene1560296}
\end{figure*}

From Figure~\ref{gene1560296}, we observe that the probes 1 and 2
are dominant probes. Further
inspection shows that the first singular vector is primarily
determined by these two probes. Following the method of Lu et al. (\citeyear{Lu}),
we find that Probes 1, 2 and 3 are remapped to three transcripts each
(``veejee.aApr07-unspliced,'' ``DST.vlApr07-unspliced'' and
``DST.iApr07''), yet the other probes are remapped to two variants only
(``veejee.aApr07-unspliced'' and ``DST.vlApr07-unspliced''). For this probe set,
the significant 2-dimensional mean structure of the data matrix
could be resolved by proper remapping of the probes.

\subsection{\texorpdfstring{Example 2.}{Example 2}}\label{example2}

In this example
the data
(\href{http://www.ncbi.nlm.nih.gov/projects/geo/query/acc.cgi?acc=GSE8874}{http://www.ncbi.nlm.nih.gov/}
\href{http://www.ncbi.nlm.nih.gov/projects/geo/query/acc.cgi?acc=GSE8874}{projects/geo/query/acc.cgi?acc=GSE8874}) were collected in a recent experiment with the $2\times
2\times 2$ factorial design, the detail of which is discussed in
Leung et al. (\citeyear{Ping}). The three factors (with two levels each) are as follows:
\begin{longlist}[(iii)]
\item[(i)] mutation: mutant or wild type (WT);
\item[(ii)]tissue: retinas or whole body;
\item[(iii)]time: 36 or 52 hours post-fertilization.
\end{longlist}

Under each condition, three Affymetrix zebrafish genome arrays are
replicated, so we have 24 arrays in total. The vector
$\underline{\hat{\mu}}$ is computed as in Example 1 by assuming that
the means in each tissue group are equal. Furthermore, we generate
two directions $\underline{a}_1$ and $\underline{a}_2$, used to
reflect the possible tissue and mutation effects, respectively. In
the study we still use $\mathit{PM}$ as the intensity measure and carry out
the singular value decomposition to get the two largest singular
values as $\hat{\lambda}_1$ and $\hat{\lambda}_2$, where
$\hat{\lambda}_1\geq\hat{\lambda}_2$. We focus on 75 probe sets with
the highest $\hat{\lambda}_2^2/\hat{\lambda}_1^2 $ (with all those
ratios above $1/10$), and use the $\chi^2$ test described in Section
\ref{chi2} on each of those probe sets.

In this example 39 out of 75 probe sets are detected as
individually significant, out of which 39 probe sets remain
significant after the multiple test adjustment of Benjamini and Hochberg (\citeyear{Benj1}). We
shall describe one such probe set in detail. 

\begin{figure*}[b]

\includegraphics{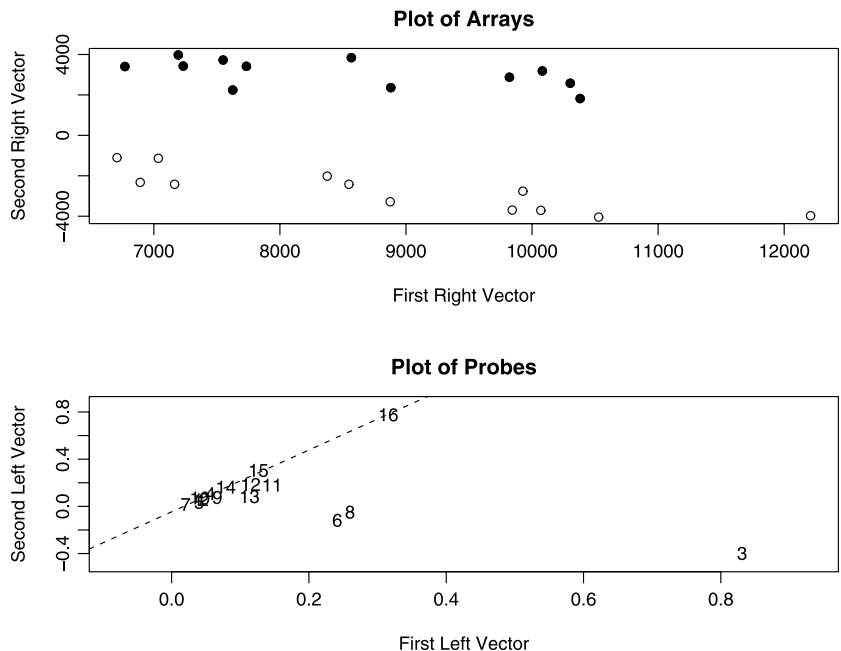}

  \caption{Scatterplot of singular vectors for the probe set ``Dr.7506.1.A1\_at.''
  The probe numbers are shown in the lower plot and the dotted line is a robust linear fit.
  The circles in the upper plot represent the arrays hybridized by the samples from retinas,
  while the solid points represent the arrays hybridized by the samples from whole body.}\label{gene585}
\end{figure*}

\subsubsection{\texorpdfstring{Probe set ``Dr.7506.1.A1\_at.''}{Probe set ``Dr.7506.1.A1\_at''}}
In the zebrafish genome array, the probe set ``Dr.7506.1.A1\_at''
corresponds to gene ``tuba8l2.'' The $\chi^2$ test gave the $p$-value
of $2.37 \times 10^{-5}$, the adjusted $p$-value of $7.52 \times
10^{-5}$ and the $q$-value of $4.83\times 10^{-6}$. The first four
singular values are (43142, 14839, 2078, 1688). It is clear from
Figure~\ref{gene585} that we cannot distinguish two tissue groups
based on $\hat{\theta}_{1i}$, but the two groups are well separated
by $\hat{\theta}_{2i}$. Further inspection of the data shows that
the intensities of Probe 3 are linearly related with
$\hat{\theta}_{1i}$,  but $\hat{\theta}_{2i}$ are linearly related
with the intensities of Probe 15. From Table~\ref{compare585}, we
see that the information from $\hat{\theta}_{2i}$ are clearly
nonnegligible. Furthermore, we used BLAST to verify that all the
probes are appropriate for Gene ``tuba8l2,'' so there is strong
evidence that the expression profile for Gene ``tuba8l2'' cannot be
summarized by the usual unidimensional index across experimental
conditions. In fact, the commonly used gene expression index would
mask the clear differential expressions of the two tissue types.

\begin{table}
\caption{A summary of the absolute values of
$\hat{\theta}_{2i}\hat{\phi}_{2j}/\hat{\theta}_{1i}\hat{\phi}_{1j}$
in percentage by probes}\label{compare585}
\begin{tabular*}{\textwidth}{@{\extracolsep{\fill}}ld{2.2}d{2.2}d{2.2}d{2.2}d{3.2}@{}}
\hline
 \textbf{Dr.7506.1.A1\_at}
 &\multicolumn{1}{c}{\textbf{Min. (\%)}}
 &\multicolumn{1}{c}{\textbf{Q1 (\%)}}
 &\multicolumn{1}{c}{\textbf{Med. (\%)}}
 &\multicolumn{1}{c}{\textbf{Q3 (\%)}}
 &\multicolumn{1}{c@{}}{\textbf{Max}. (\%)}
 \\
 \hline
Probe 1&  23.71&  40.44&  48.73&  58.58&  81.25\\
Probe 2&   20.13&  34.33&  41.37&  49.73&  68.97\\
Probe 3&   7.76&   13.23&  15.94&  19.16&  26.58\\
Probe 4&   30.74&  52.42&  63.16&  75.94&  105.32\\
Probe 5&   13.20&  22.51&  27.12&  32.60&  45.22\\
Probe 6&   7.95    &13.56& 16.33&  19.64&  27.23\\
Probe 7&   12.37   &21.10  &25.42& 30.56&  42.38\\
Probe 8&   3.08    &5.25   &6.32   &7.60&  10.54\\
Probe 9&   18.24   &31.10  &37.48  &45.05& 62.48\\
Probe 10&  27.56   &47.00  &56.63  &68.08  &94.42\\
Probe 11&  19.89   &33.92  &40.87  &49.13  &68.14\\
Probe 12&  26.07   &44.47  &53.58  &64.42  &89.34\\
Probe 13&  11.71   &19.98  &24.07  &28.94  &40.13\\
Probe 14&  33.25   &56.70  &68.32  &82.14  &113.92\\
Probe 15&  38.84   &66.24  &79.81  &95.96  &133.08\\
Probe 16&  39.66   &67.64  &81.50  &97.98  &135.88\\
\hline
\end{tabular*}
\end{table}

\section{\texorpdfstring{Conclusions.}{Conclusions}}\label{conclude}

In this article we have proposed a new framework for testing
the unidimensional
mean structure of the probe-level data matrix. For most applications, we
can carry out the tests discussed in the article based
on large sample approximations.  We also proposed a model-based
bootstrap algorithm to better control type I errors when the sample size is
small.

In two case studies, the proposed method detected genes whose
expression levels were not well summarized by unidimensional
indices. Through detailed inspection of the probe-level intensities
of those genes, we found that the intensities of different probes
can show different profiles across experimental conditions. In our investigation,
we noticed that the following scenarios exist for the violation of a unidimensional gene expression summary:
\begin{longlist}[(4)]
\item[(1)] A large percentage of probes that have poor binding strengths or low intensity measures in a probe set
can mask the gene expression profiles.
\item[(2)] One or more probes should be remapped to different variants of the same gene.
\item[(3)] One or more probes are cross-hybridized.
\item[(4)] An outlying and erroneous measurement is present for one of the probes.
\item[(5)] The multiplicative model used to summarize gene expression is inadequate even with all the probes
well selected.
\end{longlist}

It has been observed by
Harbig, Sprinkle and Enkemann (\citeyear{Harbig}) that outlier signals on just one probe can
seriously affect the calculations used for the subsequent analysis.
While we do not always have definite answers as to the biological
implications of such structures, our statistical analysis is valuable in both flagging
the potentially interesting and important probes and genes for further
scientific investigations. Our approach does not lead directly to
probe remapping, but may suggest candidates for
possible alternative mapping [Gautier et al. (\citeyear{Gautier}); Lu et al. (\citeyear{Lu})]. The bottom line is clear: if we
solely rely on models that assume unidimensional gene expressions,
we might miss some of the complexities in gene expression data analysis.
When a unidimensional model is shown to be inadequate, appropriate actions,
such as probe remapping, an alternative model or a different summarization
method [e.g., Kapur et al. (\citeyear{Kapur})], are called for.

\section*{Acknowledgments}

The authors thank Doctors Ping Ma and Sheng Zhong, as well as the Associate Editor,
for their helpful suggestions on the case studies
presented in the paper.

\begin{supplement}[id=suppA]
\stitle{Proofs of Main Results}
\slink[doi]{10.1214/09-AOAS262SUPP}
\slink[url]{http://lib.stat.cmu.edu/aoas/262/Supplement.pdf}
\sdatatype{.pdf}
\sdescription{We give a lemma on consistency, followed by the proofs for the theorems that are described in Sections~\ref{model} and~\ref{test}.}
\end{supplement}

\printaddresses

\end{document}